\documentclass[3p,times,twocolumn]{elsarticle}

\usepackage{ecrc}

\volume{00}

\firstpage{1}

\journalname{Nuclear Physics B Proceedings Supplement}

\runauth{Shaouly Bar-Shalom}

\jid{nuphbp}

\jnltitlelogo{Nuclear Physics B Proceedings Supplement}

\usepackage{amssymb}
\usepackage{epsfig}
\usepackage{graphicx}
\usepackage{epstopdf}

\usepackage[figuresright]{rotating}

\def\lsim{\:\raisebox{-0.5ex}{$\stackrel{\textstyle<}{\sim}$}\:}
\def\gsim{\:\raisebox{-0.5ex}{$\stackrel{\textstyle>}{\sim}$}\:}
\def\leff{\lcal_{\rm eff}}
\def\dm{\delta m_h^2}

\def\ti{{\rm SM}}
\def\tii{{\rm Hvy}}
\def\tiii{{\rm eff}}

\usepackage{bbm}

\newcommand{\nc}{\newcommand}
\nc{\beq}{\begin{equation}}  \nc{\eeq}{\end{equation}}
\nc{\bea}{\begin{eqnarray}}  \nc{\eea}{\end{eqnarray}}
\nc{\baa}{\begin{array}}     \nc{\eaa}{\end{array}}
\nc{\bit}{\begin{itemize}}   \nc{\eit}{\end{itemize}}
\nc{\ben}{\begin{enumerate}} \nc{\een}{\end{enumerate}}
\nc{\bce}{\begin{center}}    \nc{\ece}{\end{center}}
\nc{\bpm}{\begin{pmatrix}}   \nc{\epm}{\end{pmatrix}}
\nc{\bvt}{\begin{verbatim}}  \nc{\evt}{\end{verbatim}}
\def\half{\frac12}	

\def\to{\rightarrow}

\def\boldoverdot{\,{\raise6pt\hbox{\bf.}\!\!\!\!\>}}

\def\lcal{{\cal L}}

\def\ocal{{\cal O}}

\def\JJ{{\bf J}}

\def\VV{{\bf V}}

\usepackage{bm}
\def\chibf{{\bm\chi}}		

\def\diag{\hbox{\diag}}

\def\gev{\hbox{GeV}}
\def\tev{\hbox{TeV}}

\def\doubleundertext#1{
{\undertext{\vphantom{y}#1}}\par\nobreak\vskip-\the\baselineskip\vskip4pt%
\undertext{\hbox to 2in{}}}
\def\inbox#1{\vbox{\hrule\hbox{\vrule\kern5pt
     \vbox{\kern5pt#1\kern5pt}\kern5pt\vrule}\hrule}}
\def\sqr#1#2{{\vcenter{\hrule height.#2pt
      \hbox{\vrule width.#2pt height#1pt \kern#1pt
         \vrule width.#2pt}
      \hrule height.#2pt}}}

\def\today{\ifcase\month\or
  January\or February\or March\or April\or May\or June\or
  July\or August\or September\or October\or November\or December\fi
  \space\number\day, \number\year}
\def\pmb#1{\setbox0=\hbox{#1}%
  \kern-.025em\copy0\kern-\wd0
  \kern.05em\copy0\kern-\wd0
  \kern-.025em\raise.0433em\box0 }

\def\pmbb#1{\setbox0=\hbox{#1}%
  \kern-.02em\copy0\kern-\wd0
  \kern.04em\copy0\kern-\wd0
  \kern-.02em\raise.03464em\box0 }
\def\up#1{^{\left( #1 \right) }}
\def\inv#1{\frac1{#1}}
\def\sumprime_#1{\setbox0=\hbox{$\scriptstyle{#1}$}
  \setbox2=\hbox{$\displaystyle{\sum}$}
  \setbox4=\hbox{${}'\mathsurround=0pt$}
  \dimen0=.5\wd0 \advance\dimen0 by-.5\wd2
  \ifdim\dimen0>0pt
  \ifdim\dimen0>\wd4 \kern\wd4 \else\kern\dimen0\fi\fi
\mathop{{\sum}'}_{\kern-\wd4 #1}}
\begin{document}

\begin{frontmatter}




\title{Softening Higgs Naturalness - an EFT Analysis}


\author[sbs]{Shaouly Bar-Shalom\corref{cor1}}
\ead{shaouly@physics.technion.ac.il}
\address[sbs]{Physics Department, Technion - Institute of Technology, Haifa 32000, Israel
 \\
$~$\\
$~$\\
$~$\\
{\it \bf Talk given at the 37th International Conference on High Energy Physics (ICHEP 2014) \\
2-9 July 2014, Valencia, Spain}
}
\cortext[cor1]{Speaker. Co-authors: Amarjit Soni and Jose Wudka}

\begin{abstract}
We investigate naturalness in the Standard Model (SM) Higgs sector
using effective field theory (EFT) techniques and find
the requirements on the new heavy physics that can potentially
cure the little hierarchy problem below a
scale $ \Lambda \gg {\cal O}(1~{\rm TeV})$, assuming
the new heavy particles have a mass
{\em larger} than $ \Lambda $.
In particular, we determine the
conditions under which the 1-loop corrections to $ m_h $  from the heavy new physics
can balance those created by SM loop effects up to the naturalness scale
$\Lambda$, a condition we denote by ``EFT Naturalness''.
We obtain the higher dimensional ($n \ge 5$) operators in the
effective Lagrangian that can lead to EFT Naturalness, and
classify the underlying heavy
theories that can generate such operators at tree-level.
We also address the experimental constraints on our EFT Naturalness
setup and discuss the expected experimental signals
of the new heavy physics associated with EFT Naturalness.
\end{abstract}




\end{frontmatter}


\section{Introduction: naturalness in the SM} \label{intro}
The recent LHC discovery of a light 125 GeV scalar particle \cite{LHCHiggs}
exacerbates the long-standing fundamental difficulty of the SM
known as the hierarchy problem. Simply put, the presence of a fundamental
Higgs with an EW-scale mass appears unnatural, since if the only physics
present up to some high scale $\Lambda$ is the SM, then it is hard
to see why the Higgs boson mass, $m_h $, does not receive large corrections of
$O(\Lambda)$. This technical difficulty is also known as the naturalness or
fine-tuning problem of the SM. The ``master equation" for naturalness
in the SM Higgs sector, which has been the driving force behind
the search for new physics (NP) in the past several decades,
is the leading
$O(\Lambda^2)$ 1-loop
corrections to the SM Higgs mass squared:
\begin{eqnarray}
\dm(\ti) = \frac{\Lambda^2}{16 \pi^2}
\left[ 24 x_t^2 -6 \left(2 x_W^2 + x_Z^2  + x_h^2 \right) \right]
~, \label{deltaSM}
\end{eqnarray}
where $x_i \equiv \frac{m_i}{\rm v} ~~ ({\rm v} \simeq 246 ~{\rm GeV})$, so that
the dominant contribution is
generated by the top-quark loop.
In particular, The
hierarchy/naturalness problem of the SM corresponds to the fact
that $ \dm(\ti) > m_h^2({\rm tree}) $ when $ \Lambda \gtrsim 500~\gev $,
assuming $m_h ({\rm tree}) $ is close to the observed value, i.e.,
$m_h({\rm tree}) \simeq m_h\simeq 125~\gev $.

The conventional way to address this problem is to postulate
the existence of NP beyond the SM, involving new heavy particles
and possibly new symmetries, which soften or exactly cancel
the $O(\Lambda^2)$ divergences in Eq.~\ref{deltaSM}.
One commonly used approach is to assume that the SM is not
a complete description of the heavy physics {\it below} $ \Lambda$,
there being other particles yet to be discovered with masses below this scale.
In this case, the parameters of the theory are such that there are
cancellations between the SM and NP contributions to $\dm $.
This is what happens e.g., in supersymmetric and in Little-Higgs models, or in
simpler model-dependent or phenomenological extensions of the SM's scalar
and/or fermion sector (see, {\it e.g.} \cite{jose-pragmatic,jose-DM-singlets,craig}).

Our approach - ``EFT Naturalness" - is different and has a modest goal \cite{our-nat-paper}:
we wish to acquire insight
regarding the underlying NP which can potentially cure only the
{\it little hierarchy} problem of the SM up to some high scale $ \Lambda \gg {\cal O}(1~{\rm TeV})$,
so we will not be concerned with the issues pertaining to the UV-completion.

\begin{figure}[here]
\begin{center}
\includegraphics[scale=0.3]{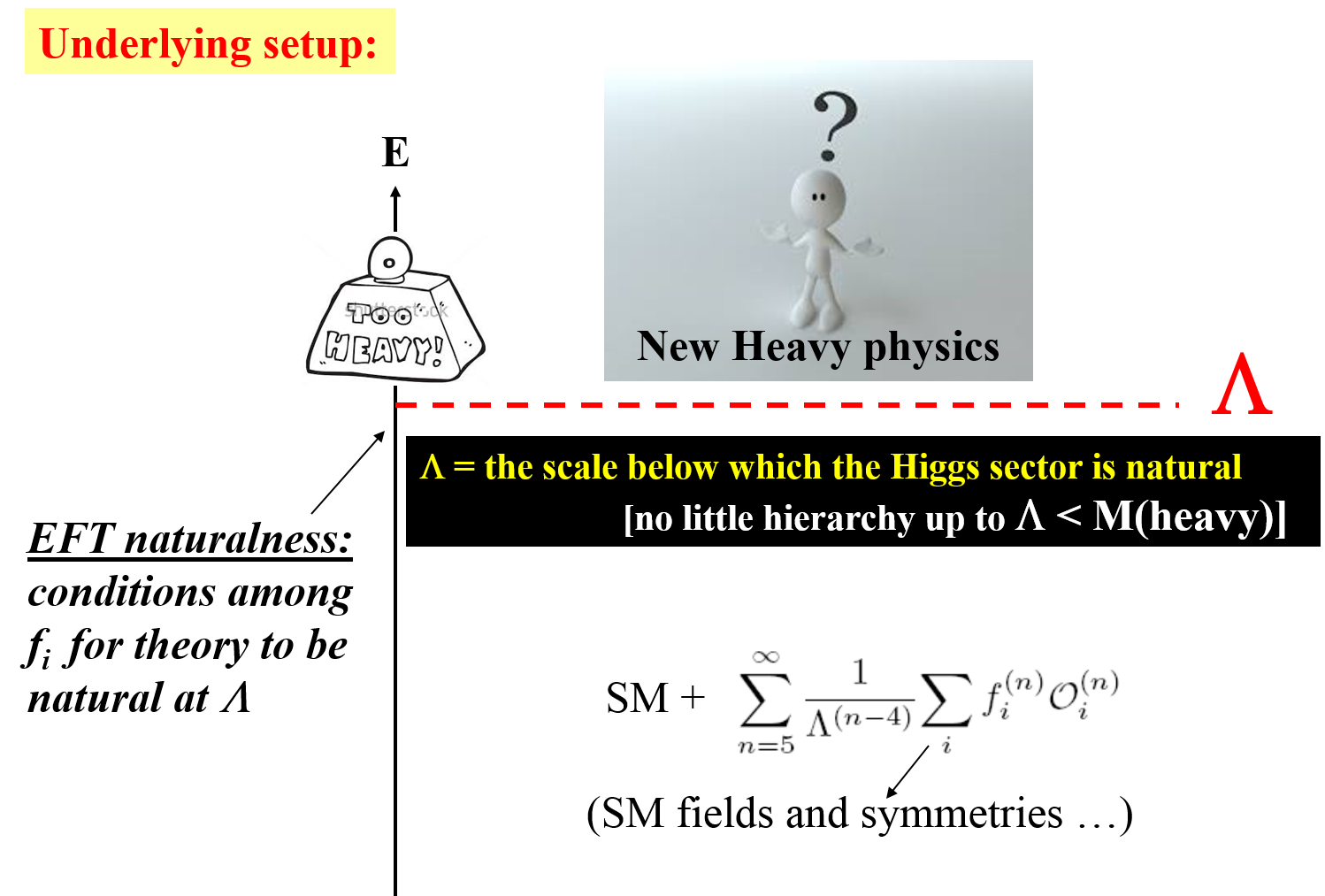}
\caption{\emph{The ``EFT Naturalness" setup.}
\label{fig:EFTnat}}
\end{center}
\end{figure}

Our EFT Naturalness prescription is thus defined as follows (also
summarized in Fig.~\ref{fig:EFTnat}):
\begin{itemize}
\item We assume that the underlying
NP which can potentially restore naturalness to the Higgs sector
lies {\it above} $ \Lambda$.
\item We, therefore, exploit EFT techniques to
parameterize the heavy NP {\it above} $ \Lambda$.
\item We define $\Lambda$ to be the scale
below which naturalness in the Higgs sector can be restored,
i.e., no little hierarchy up to $\Lambda < M$, where $M$
represents the typical mass scale of the new heavy particles.
\item We look for the EFT Naturalness conditions:
the conditions on the physics which lies above
$\Lambda$ that can soften naturalness in the Higgs sector, i.e.,
can cure the little hierarchy problem up to $\Lambda$.
Specifically, we find the conditions under which
$ \dm(\ti) + \dm(\tiii) \lesssim m_h^2 $ when $ m_h \ll \Lambda \le M$
($M$ being the typical mass scale of the new heavy physics which lies at or
above $\Lambda$) and $\dm(\tiii)$ is the 1-loop contributions to the Higgs mass,
which are generated by the effective Lagrangian that parameterizes the heavy NP, as will
be outlined below.
\end{itemize}

\section{Guidelines for EFT}
Physics at $E \leq \Lambda$ is assumed to be described by
${\cal L}={\cal L}_{\rm SM} + \leff$, where:
\begin{eqnarray}
\leff = \sum_{n=5}^\infty \frac1{\Lambda^{(n-4)}}  {\sum_i
f_i^{(n)}  \ocal_i^{(n)}}\label{eff1} ~.
\end{eqnarray}
and $\ocal_i\up n$  are higher dimensional operators
($n$ denotes the dimension and $i$  all other
distinguishing labels), which are local, gauge and Lorentz
invariant combinations of SM fields and their derivatives.
They result from integrating out the heavy
degrees of freedom  of the heavy NP theory
that underlies the SM, and expanding in inverse powers
of $\Lambda$ after appropriate renormalization of the SM parameters.

In particular,
\begin{itemize}
\item The ``light fields" (i.e., at energies $E \leq \Lambda$) are the SM fields.
\item The guage-symmetry at $E \leq \Lambda$ is taken to be the SM
SU(3)$\times$SU(2)$\times$U(1) gauge symmetry. This is useful for
classifying the higher dimensional effective operators \cite{patterns,Wudka-new}.
\item The underlying NP is assumed to be weakly coupled, renormalizable, obeys
gauge-invariance and preserves symmetries of the known dynamics (SM).
This is also useful for classifying the higher dimensional effective operators,
see e.g., \cite{trott1}.
\end{itemize}

Three more guiding principles can further simplify/reduce the number
of relevant operators (we do not know the precise form of the
underlying heavy physics and this leads to
ambiguities in selecting the relevant operators for a given process) \cite{patterns,Wudka-new}:
\begin{enumerate}
\item The use of equations of motion, also known as the ``equivalence theorem", which
is useful when applied to strictly observable quantities. It is therefore not
practical for our naturalness study.
\item Integration by parts to dismiss surface terms.
\item Loop Classification of operators (refers to the way they can be generated in the underlying heavy theory):
loop-generated (LG) operators versus potentially tree-level generated (PTG) operators.
\end{enumerate}

As it turns out, many operators can be constructed under these conditions,
in particular $O(50)$ operators just at dimension 6 \cite{1008.4884}.
However, as we will show below, only very few
can balance the SM’s 1-loop quadratic terms.

\section{EFT and the one-loop Higgs mass corrections}

\begin{figure}[htb]
\includegraphics[scale=0.3]{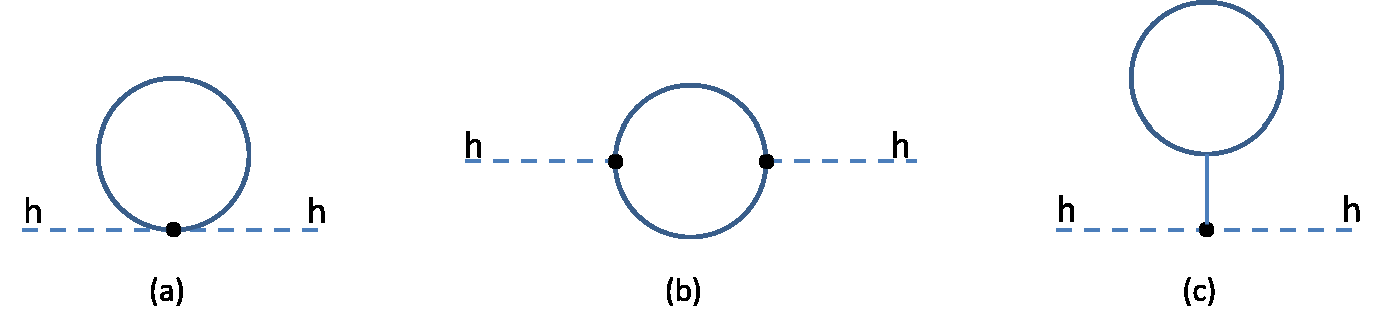}
\caption{\emph{The 1-loop graphs generating $ \dm $. The internal
lines represent bosons or fermions from either the SM or the heavy NP.}
\label{fig:graphs}}
\end{figure}
In general, {\em all} (SM and NP) one-loop corrections to $ m_h^2 $
are generated by the graphs in Fig.~\ref{fig:graphs}.
In the scenarios we are interested in here,
these corrections can be separated into 3 categories:
\begin{description}
\item{\boldmath{$\dm(\ti)$}:} When all internal lines are the light SM fields. The contributions from this category are
given in Eq.~\ref{deltaSM}.
\item{\boldmath{$\dm(\tii)$}:} When all the internal lines are heavy fields of the underlying NP.
The contributions from this category are
contained in the renormalization of the parameters of the SM that follows upon integration of the heavy particles. This is included in what we denote here as ``tree-level'' parameters, i.e., $m_h^2({\rm tree}) = m_h^2({\rm bare}) + \dm(\tii)$ (note that the tree-level mass parameter for $E < \Lambda$ within the EFT prescription need not be the same as the tree-level mass in the full theory, i.e., for $E > \Lambda$).
\item{\boldmath{$\dm(\tiii)$}:} When one line is heavy and the other is light (in graphs (b) and (c) in Fig.~\ref{fig:graphs}).
The contributions in this category are generated by the effective Lagrangian in Eq.~\ref{eff1}
and are the ones we are interested in here.\footnote{It is important to note that Eq.~\ref{eff1} can be used to
calculate such NP effects provided all energies (including those that appear within loop
calculations) are kept below $ \Lambda $.}
\end{description}

Specifically,  our EFT Naturalness approach
corresponds to finding those effective interactions
that can tame the little hierarchy problem, i.e,
leading to
$ \dm(\ti) + \dm(\tiii) \lesssim m_h^2 $ when $ m_h \ll \Lambda \le M$,
$M$ being the typical mass scale of the new heavy physics which lies at or
above $\Lambda$.

\begin{figure}[htb]
\includegraphics[scale=0.3]{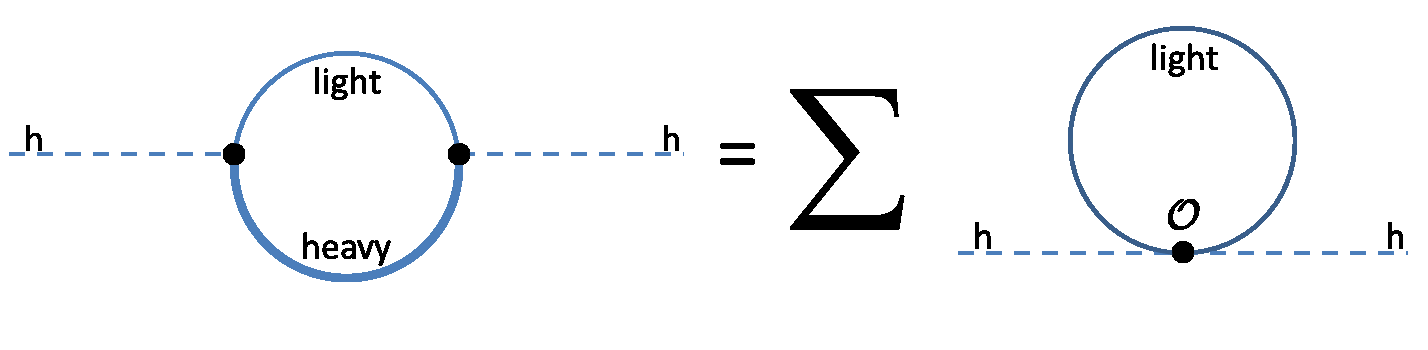}
\caption{\emph{Description of the manner in which the effective
Lagrangian Eq.~\ref{eff1} generates graphs in category $\dm(\tiii)$
defined in the text. } \label{fig:expansion}}
\end{figure}

Integrating out the heavy fields with
mass $M$, one generates an infinite series
of vertices suppressed by inverse powers of $\Lambda$ (recall $\Lambda < M$). This is
schematically depicted in Fig.~\ref{fig:expansion}, where the effective 4-particle contact
vertices on the right hand side correspond
to those  generated by the effective operators in Eq.~\ref{eff1}.
Thus, to calculate $\dm(\tiii)$, we need the set of operators $ \ocal $ which are
not LG and which give the leading contribution to the 1-loop diagrams in
Fig.~\ref{fig:graphs}(a), where the vertex is generated by the effective operator.

\begin{figure}[htb]
\includegraphics[scale=0.3]{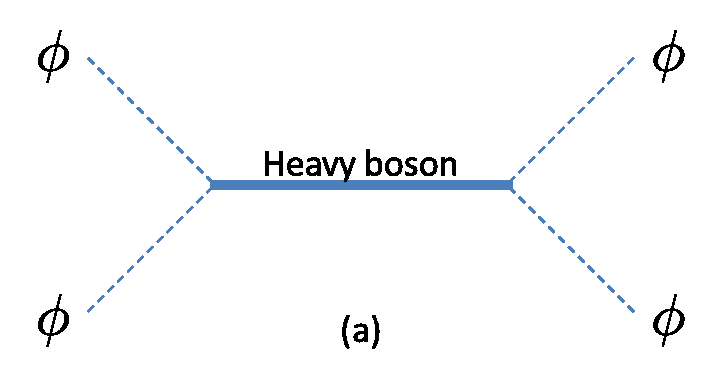} \quad
\includegraphics[scale=0.3]{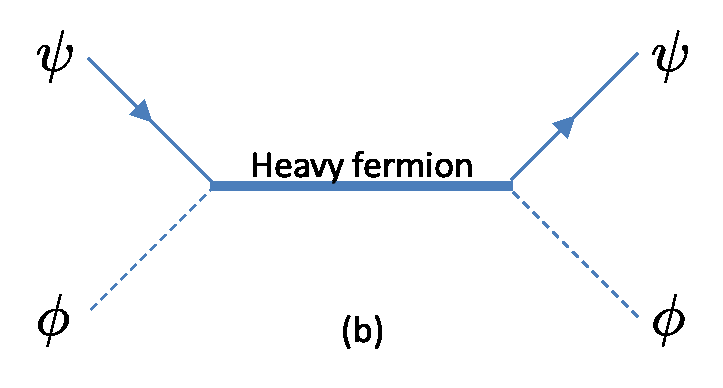} \quad
\includegraphics[scale=0.3]{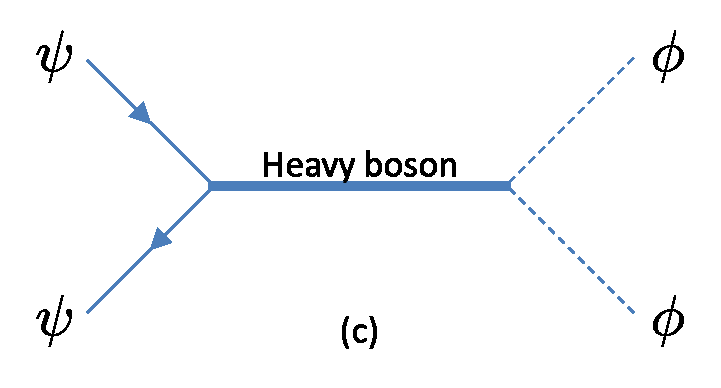}
\caption{\emph{Tree-level graphs that generate the effective operators
of type I (diagram a) and II (diagrams b and c),
that can produce leading corrections to $ \dm $. $ \phi $ and $ \psi $ denote
the SM scalar doublet and fermions, respectively and all vertices
are understood to be invariant under SM gauge transformations.}
\label{fig:np}}
\end{figure}

In \cite{our-nat-paper} we have outlined the arguments used to identify all the
higher dimensional effective operators that can generate the desired 1-loop $O(\Lambda^2) $
contributions to $ \dm$, and found that they can be
of {\it only} two types - both types can be derived by integrating out
the heavy fields exchanged in the tree-level diagrams depicted
in Fig.~\ref{fig:np}:\footnote{Note that the graphs in Fig.~\ref{fig:np} represent the possible types of NP that
can generate the effective operators in
Eqs.~\ref{eq:s-ops}, \ref{eq:v-ops} and \ref{eq:f-ops} at tree-level. There are other
types of NP that can also generate these operators, but only via loop
diagrams. It then follows that the coefficients of the
operators associated with the same heavy particle are correlated,
see discussion in \cite{our-nat-paper}.}
\begin{itemize}
\item Type I: $ \ocal $  contains 4 scalar fields, any number of derivatives and is not LG.
\item Type II: $ \ocal $  contains 2 fermions and 2 scalar fields, any number of derivatives and is not LG.
\end{itemize}

The operators of type I can be further subdivided into those that
are generated by tree-level heavy
scalar or heavy vector exchanges in diagram Fig.~\ref{fig:np}(a).
The heavy scalar exchanges gives
\begin{eqnarray}
\ocal_S\up{2k+4} &=& \half |\phi|^2 \Box^k |\phi|^2~, \nonumber \\
\ocal_{\chibf}\up{2k+4} &=& \half (\phi^\dagger \tau_I \phi) D^{2k}
(\phi^\dagger \tau_I \phi)~, \nonumber \\
\ocal_{\tilde\chibf}\up{2k+4} &=& \inv4 (\phi^\dagger \tau_I \tilde\phi) D^{2k}
(\tilde\phi^\dagger \tau_I \phi)
\label{eq:s-ops} ~,
\end{eqnarray}
which correspond to the cases where the heavy scalar is a SM gauge singlet (labeled $S$)
or an isotriplet of hypercharge 0 or 1 (labeled $ \chibf$ and $ \tilde \chibf$, respectively).\footnote{There are no other scalar operators of this type since $S,~\chibf$ and $ \tilde \chibf$ are the only
possible three states
that can be formed with two SM scalar isodoublets.}
In the following we denote these
heavy scalars collectively by  $ \Phi $.

Similarly the operators generated by heavy vector exchanges in Fig.~\ref{fig:np}(a) are
\begin{eqnarray}
&& \ocal_v\up{2k+6} = \half j _\mu \Box^k j ^\mu ~; ~
j^\mu \equiv i \phi^\dagger  D^\mu \phi + {\rm H.c.} ~, \nonumber \\
&& \ocal_{\tilde v}\up{2k+6} =
\tilde j^\dagger_\mu \Box^k \tilde j^\mu ~; ~
\tilde j^\mu \equiv i \tilde\phi^\dagger  D^\mu \phi ~, \label{eq:v-ops} \\
&&\ocal_{\VV}\up{2k+6} = \inv6 J_{I \mu} D^{2k} J^\mu_I ~;~
 J^\mu_I \equiv i \phi^\dagger \tau^I D^\mu \phi + {\rm H.c.}
~, \nonumber
\end{eqnarray}
where the labels in Eq.~\ref{eq:v-ops} refer to heavy vector isosinglets ($v,\,\tilde v$) of hypercharge 0 or 1, respectively,
and a heavy vector isotriplet ($\VV$) of hypercharge 0.
In the following we will collectively denote these heavy
vectors by $X$.

Finally, the type II operators are generated in the underlying heavy
theory by the graph in Fig.~\ref{fig:np}(b), which involves
 an exchange of a heavy fermion $ \Psi $ that may or may not be colored
 and has the same quantum numbers as
$ \phi \psi $ or $ \tilde \phi \psi $. That is, $\Psi$ can be
an isosinglet, doublet
or triplet heavy lepton or quark of hypercharge $ y_\Psi = y_\psi \pm1/2 $ ($y_r$ denotes
the hypercharge of $r$). These $\Psi$-generated operators  are
\beq
\ocal\up{2k+4}_{\Psi -\psi}=
|\phi|^2\,  \bar\psi \left(i \not\!\!D\right)^{2k-1} \psi, \quad (k\ge1)
\label{eq:f-ops}~,
\eeq
where $ \psi $ is any SM fermion.\footnote{Another type of operator that may be generated by the heavy-fermion exchange is
$( \phi^\dagger \tau_I \phi ) \, (\bar\psi \tau_I\not\!\!D^{2k-1} \psi)$,
where $ \psi $ is an isodoublet. However, this operator
will yield a contribution to $ \dm$ which is
suppressed by a factor of $ m_\psi^2/\Lambda^2 $ and is, therefore, subdominant.}

Calculating now the 1-loop quadratic corrections to $m_h^2$ which are generated by the operators in
Eqs.~\ref{eq:s-ops}, \ref{eq:v-ops} and \ref{eq:f-ops}, we obtain
\cite{our-nat-paper}:
\beq
\dm(\tiii) = -\frac{\Lambda^2}{16 \pi^2} F\up{\rm eff} ~,
\label{deltaeff}
\eeq
where
\begin{eqnarray}
F\up{\rm eff} &=&
\sum_{k=0}^\infty  \frac{1}{k+1}  \sum_\Phi f_\Phi\up{2k+4} \nonumber \\
&-& \sum_{k=0}^\infty  \frac{1}{k+2}  \sum_X f_X\up{2k+6} \nonumber \\
&-& \sum_{k=1}^\infty \frac{(-1)^k}{k+1}  \sum_{\Psi,\psi} f_{\Psi-\psi}\up{2k+4}
 ~. \label{dmh.eff}
\end{eqnarray}
with $\Phi= S,~\chibf,~\tilde \chibf$ and $ X = v,~\tilde v,~\VV$.

\section{The ``road map" to EFT Naturalness}

Let us define the measure for fine-tuning to be $\Delta_h \equiv |\dm|/m_h^2$,
where $\dm =  \dm(\ti) + \dm(\tiii) $
and $m_h^2$ is the physical mass, $ m_h^2 = m_h^2({\rm tree}) + \dm $.
We then have:
\begin{eqnarray}
\Delta_h = \frac{\Lambda^2}{16 \pi^2 m_h^2} \left| F\up{\rm eff} - 8.2 \right|
\label{ft}~,
\end{eqnarray}
using $24 x_t^2 -6 \left(2 x_W^2 + x_Z^2  + x_h^2 \right) \sim 8.2$ in Eq.~\ref{deltaSM}.

Re-writing the above defined fine-tuning condition
as $ | m_h^2({\rm tree})/\dm +1| =1/\Delta_h$, we see that
a cancellation  must occur to a precision of $ 1/\Delta_h $,
so that a larger $\Delta_h$ corresponds to a less natural
theory. Inspection of Eq.~\ref{ft} shows that, for naturalness to be restored
at $\Lambda$, the cancellation must occur among
the 1-loop contributions (i.e., leading to a natural
$m_h({\rm tree}) \sim m_W$) or else there should be a
correlation between $m_h({\rm tree})$ and $\delta m_h$, in which case
$m_h({\rm tree}) \sim m_W$ will require fine-tuning.

\begin{figure}[htb]
\includegraphics[scale=0.5]{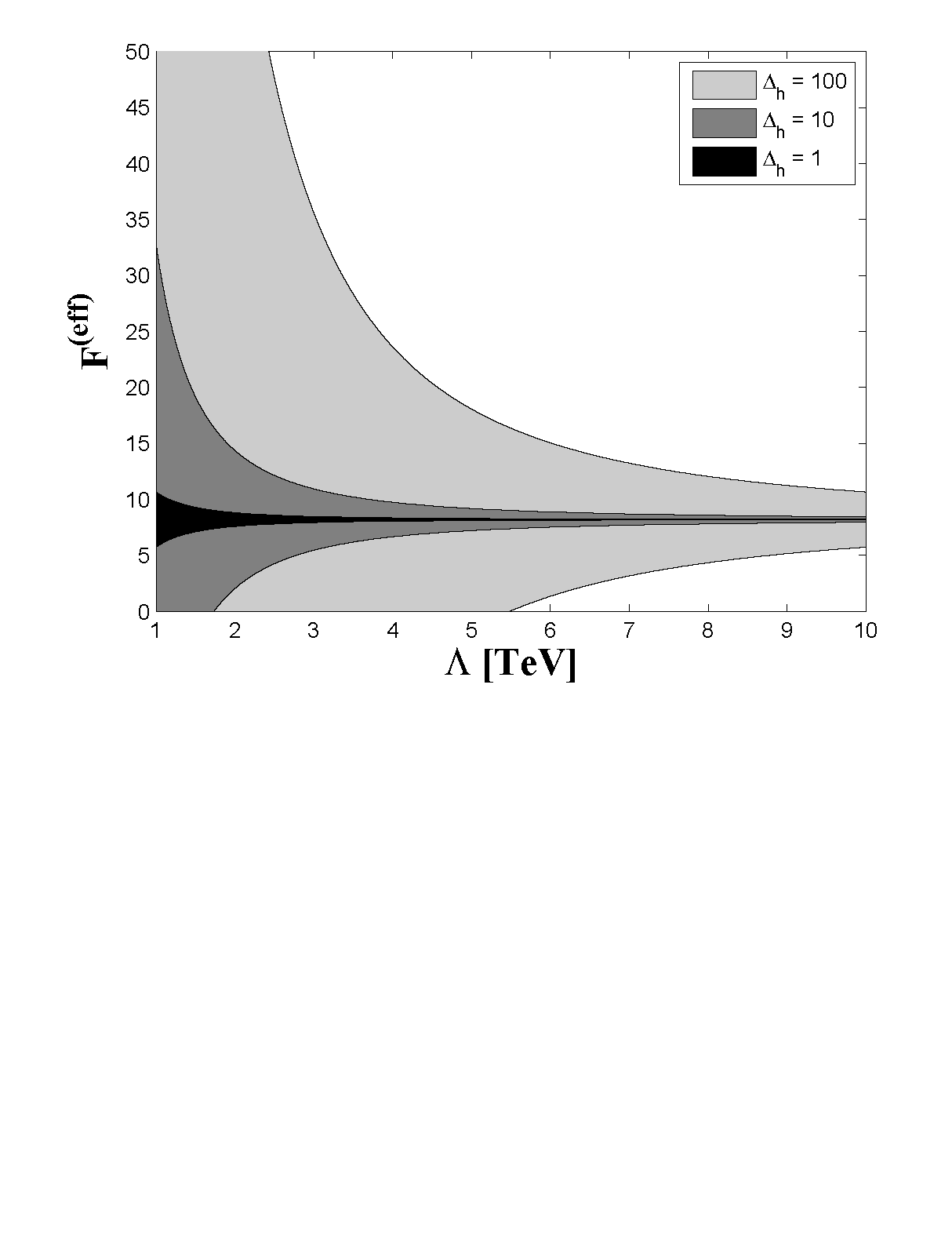}
\vspace{-4.8cm}
\caption{\emph{Regions in the $F\up{\rm eff}-\Lambda$ plane
where naturalness can be restored with no fine-tuning
($\Delta_h = \dm/m_h^2 = 1$, in black) and with fine-tuning at the level
of 10\% (dark gray) and 1\% (light gray), corresponding to
$\Delta_h = \dm/m_h^2 =10$ and 100, respectively. See also text.}
\label{fig3}}
\end{figure}

Therefore, a theory (i.e., $F\up{\rm eff}$) for which $\Delta_h=1$ is natural,
while one with $\Delta_h=10(100)$ suffers from fine-tuning of (no worse than) 10\%(1\%).
In Fig.~\ref{fig3} we plot regions in the $F\up{\rm eff} - \Lambda$ plane
that correspond to NP theories which are natural
(i.e., enclosed within the $\Delta_h = 1$ region) and
those that suffer from fine-tuning of no worse than 10\% and 1\%,
corresponding to $ \Delta_h=10$ and $\Delta_h=100$, respectively.
We find, for example, that theories for which
$8.17 \lesssim F\up{\rm eff}  \lesssim 8.23$
are natural at $\Lambda \sim 10$ TeV, while theories with
$7.95 \lesssim F\up{\rm eff}  \lesssim 8.45$ or
$5.73 \lesssim F\up{\rm eff}  \lesssim 10.67$ will suffer from
10\% or 1\% fine-tuning, respectively, at $\Lambda \sim 10$ TeV.
Clearly, if the NP scale for EFT Naturalness
is $\Lambda \sim 5$ TeV, then a much wider range of theories
are allowed if one is willing to tolerate 1\% fine-tuning, in particular
those giving $0 \lsim F\up{\rm eff}  \lsim 18$.\footnote{It should also be
noted that the EFT Naturalness regions
shown in Fig.~\ref{fig3} may in general be subject to additional constraints
(e.g., from perturbativity),
depending on the details of the specific underlying theory.}

Finally, we emphasize that, for a specific heavy NP model, $F\up{\rm eff}$ will be a dimensionless function of the
NP parameters and $\Lambda$, such as the ratio $ \Lambda/M$ (where $M$ denotes one of the heavy particle masses, see next section).
Thus, $F\up{\rm eff}$ is in general cutoff dependent, so that the cancellation conditions will depend on
$ \Lambda$ as well. Our requirement that $ \Lambda $ be the scale below which the SM little-hierarchy
problem is solved is consistent within our EFT Naturalness scenario because of the requirement $ M>\Lambda $.

\section{Less ignorance: insight for a natural heavy NP}

Let us now examine the general content and properties of the heavy NP
that can generate the effective operators in
Eqs.~\ref{eq:s-ops}, \ref{eq:v-ops} and \ref{eq:f-ops} at tree-level,
and find the EFT Naturalness conditions within these theories.
In particular, we can calculate the cutoff-dependent
coefficients $f_\Phi\up{2k+4}$, $f_X\up{2k+6}$ and $f_{\Psi-\psi}\up{2k+4}$ in Eq.~\ref{dmh.eff},
in terms of the parameters (masses and couplings) of the underlying NP theory.

The general form of the potentially natural NP which contains
the relevant interactions of the heavy scalars
($S,~\chibf,~\tilde\chibf$) to $\phi^2$, of the heavy vectors
($ v,~\tilde v,~\VV$) and of the heavy fermions
($\Psi$) to $\psi \phi$ are \cite{our-nat-paper}:
\bea
\Delta \lcal_{\Phi} &=&  u_S  S|\phi|^2 + u_\chibf  \phi^\dagger \chibf \phi
+ \half \left( u_{\tilde\chibf} \tilde\phi^\dagger \tilde\chibf \phi + {\rm H.c} \right) ~, \cr
\Delta \lcal_{\Psi} &=& \sum_{\Psi,\,\psi} \left( y_{\Psi-\psi} \bar\psi \Psi \phi + {\rm H.c} \right) ~, \cr
\Delta\lcal_X &=&  g_v v_\mu j^\mu
+ g_\VV \VV_\mu \cdot \JJ^\mu
+ \left(g_{\tilde v} {\tilde v}_\mu {\tilde j}^\mu + {\rm H.c.} \right) ~,
\label{sr}
\eea
where the currents $j,~\tilde j$ and \JJ\ (with components $J_I$) are defined in Eq.~\ref{eq:v-ops}.

Using Eq.~\ref{sr} to calculate the diagrams in Fig.~\ref{fig:np}
in the limit that the masses are larger than the typical momentum transfer involved,
we find:\footnote{Note that the expressions in Eq.~\ref{eq:fs}
hold only when $M_{\Phi,\Psi,X} \gtrsim\Lambda$.}
\begin{eqnarray}
f_\Phi\up{2k+4}(\Lambda) &=& \left| \frac{u_\Phi}{M_\Phi} \right|^2 \left( \frac{-\Lambda^2}{M_\Phi^2}\right)^k ~, \nonumber \\
f_{\Psi-\psi}(\Lambda) \up{2k+4} &=& \half I_\Psi |y_{\Psi-\psi} |^2 \left( \frac{\Lambda^2}{M_\Psi^2} \right)^k ~, \nonumber \\
f_X\up{2k+6}(\Lambda) &=& I_X |g_X|^2  \left( \frac{-\Lambda^2}{M_X^2}\right)^{k+1} ~,
\label{eq:fs}
\end{eqnarray}
where $ I_\zeta=1,\,2,\,3$ when the field $\zeta$ is an isosinglet, doublet or triplet,
respectively, and $M_{\Phi,\Psi,X} \gtrsim\Lambda$ are the masses of the heavy scalars,
fermions and vectors, respectively.

Thus, matching the effective theory at $\Lambda$ and
using the definition of $F\up{\rm eff}$ in Eq.~\ref{dmh.eff},
we obtain the 1-loop corrections to the Higgs mass in terms of the parameters
of the new heavy physics
(i.e., from
$\Delta {\cal L}_{NP}=\Delta {\cal L}_{\Phi}+\Delta {\cal L}_{\Psi} + \Delta {\cal L}_{X}$):
\begin{eqnarray}
F\up{\rm eff}(\Lambda) &=&
\sum_\Phi \frac{|u_\Phi|^2}{M_\Phi^2} A\left( \frac{\Lambda^2}{M_\Phi^2} \right) \nonumber \\
&+& \half \sum_{\Psi,\psi} I_\Psi |y_{\Psi-\psi}|^2  \left[1-
A\left( \frac{\Lambda^2}{M_\Psi^2}\right) \right] \nonumber \\
&+& \sum_X I_X |g_X|^2  \left[1-
A\left( \frac{\Lambda^2}{M_X^2}\right) \right] \label{Feffmodels}~,
\end{eqnarray}
where $A(x)=\ln(1+x)/x$, so that
$1>A(x) \ge0 $, from which it follows that $ F\up{\rm eff} > 0 $.

The upshot of this phenomenological study of EFT Naturalness is that,
given the masses and couplings of the new heavy states (heavy fermions, scalars and/or vectors), we
can derive the scale $\Lambda$ - below which the corresponding  underlying
theory is natural or has a certain degree of
fine-tuning. In particular, the EFT Naturalness scale can be written as
$\Lambda=\Lambda(\Delta_h,g_{NP},M_{NP})$,
where $g_{NP}$ and
$M_{NP} > \Lambda$ are the couplings
and masses of the new heavy states.
Naturality at $\Lambda$ (i.e., below which
the corresponding extension to the SM is natural) is, therefore, given by
$\Lambda = \Lambda(\Delta_h=1,g_{NP},M_{NP})$.

To illustrate the above, let us consider the simplest model for EFT Naturalness, where
the SM is minimally extended by one real scalar singlet $S$, with a mass $M_S$ and a super-renormalizable scalar coupling $u_S$ in
Eq.~\ref{sr}, i.e., the case
where ${\cal L} = {\cal L}_{SM} + u_S  S|\phi|^2$. In particular, we wish to find
the scale $\Lambda$
below which this one singlet extension of the SM is natural?
As we have shown above, this scale will depend on the
mass ($M_S$) and coupling ($u_S$) of the singlet $S$ to
$\phi^2$, i.e., $\Lambda = \Lambda(1,u_S,M_{S})$.

From Eq.~\ref{Feffmodels} we can obtain the 1-loop contribution of a heavy scalar singlet to the Higgs mass:
\beq
\dm(\tiii;S) = -\frac{|u_S|^2}{16 \pi^2}
\ln\left(1+ \frac{\Lambda^2}{M_S^2} \right)~,
\eeq
so that the overall 1-loop correction to the Higgs mass in this model is (keeping only the top-quark loop for the SM contribution):
\begin{eqnarray}
\dm &=& \dm({\rm SM}) + \dm(\tiii;S) \nonumber \\
&\approx& \frac{\Lambda^2}{16 \pi^2} \cdot \left( 24 \frac{m_t^2}{v^2} -\frac{|u_S|^2}{\Lambda^2}
\ln\left(1+ \frac{\Lambda^2}{M_S^2} \right) \right) ~.
\end{eqnarray}

\begin{figure}[htb]
\includegraphics[scale=0.3]{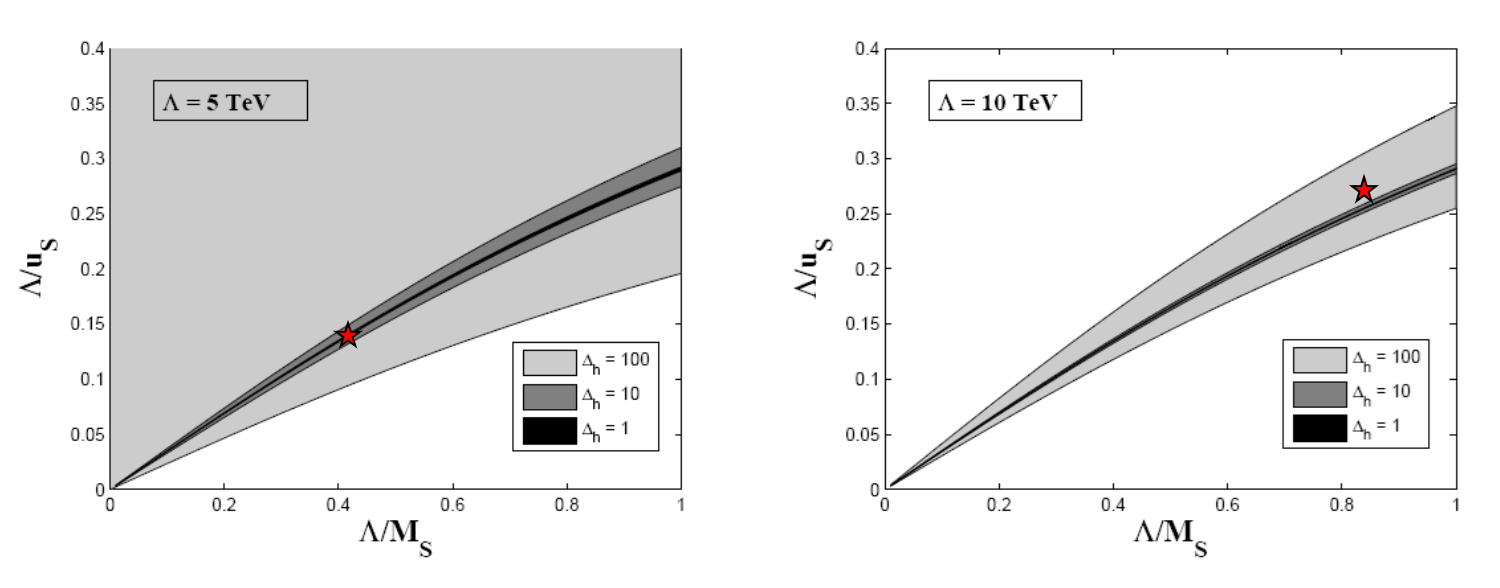}
\vspace{-0.3cm}
\caption{\emph{Regions in the $({\Lambda}/{u_S}) - ({\Lambda}/{M_S})$ plane,
for $\Lambda=5$ TeV (left) and $\Lambda=10$ TeV (right),
where naturalness can be restored with no fine-tuning
($\Delta_h = \dm/m_h^2 = 1$, in black) and with fine-tuning at the level
of 10\% (dark gray) and 1\% (light gray), corresponding to
$\Delta_h = \dm/m_h^2 =10$ and 100, respectively. The red
stars indicate the case where the scalar singlet has a mass
$M_S=12$ TeV and a super-renormalizable coupling $u_S=36$ TeV.}
\label{figsinglet}}
\end{figure}

In Fig.~\ref{figsinglet} we plot the regions in the
$\Lambda/u_S - \Lambda/M_S$
plane which correspond to $\Delta_h = 1,~10$ and 100, for a fixed NP naturalness
scale
of either $\Lambda=5$ TeV or $\Lambda =10$ TeV. Note that
$0 < \Lambda/M_S \leq 1$ (i.e., $M_S \geq \Lambda$ for the EFT
prescription to be valid, see discussion above).
Thus, we can find from Fig.~\ref{figsinglet} the values of ($M_S,u_S$)
for which the one scalar singlet extension of the SM can restore naturalness
in the Higgs sector up to $\Lambda=5$ TeV or $\Lambda =10$ TeV.
For example, as marked on the plot, if
$M_S \sim 12$ TeV and $u_S \sim 36$ TeV, then the Higgs
sector of this model is natural up to $\Lambda=5$ TeV,
while it will require fine-tuning at the level of
$ \sim 4\%$ for $\Lambda \sim 10$ TeV, i.e., if no additional
new physics appear below $\sim 10$ TeV.
This result should be interpreted as follows:
the one scalar singlet extension of the SM with this choice of parameters (i.e., $M_S \sim 12$ TeV and $u_S \sim 36$ TeV) is natural up to scales
$ \sim 5 ~ \tev $ (an order of magnitude improvement over the pure SM).

\section{Constraints from current data}

In \cite{our-nat-paper} we have studied the constraints that the
present data imposes on our EFT Naturalness operators.
We found that these operators can cause 2 types of effects that
need to be considered:
a shift to the $\rho$-parameter and a shift
of the SM Higgs couplings to the SM fermions and gauge-bosons, which, therefore, needs to be confronted with the measured
production and decay rates of the recently discovered 125 GeV Higgs state.

We have found that while in the latter case (i.e., the deviations of the Higgs production and decay rates) no useful limit can be
obtained, the well measured value of the $\rho$-parameter
imposes a bound of $\Lambda \gsim 10$ TeV on the scale of the operators involving the heavy scalar triplets and heavy vector exchanges.

\section{Collider signals of EFT Naturalness}
Let us briefly discuss the potential collider signals of our
EFT Naturalness operators, or equivalently, of the NP that can
restore naturalness at energy scales which are accessible
to current and future high energy colliders.

As for the heavy singlet example discussed above, it is expected to have a small
mixing with the CP-even component of the SM Higgs doublet (see e.g., \cite{1407.5342,singletlimit}) and is,
therefore, unlikely to be detected at the 14 TeV LHC; whether it is produced
through Higgs bremshtralung $h_{SM}^\star \to h_{SM} S$ or in the s-channel, leading
e.g., to $pp \to S \to h_{SM} h_{SM}$ \cite{1310.6035}.

On the other hand, in the more general case (where the heavy new physics includes
new heavy triplet bosons and/or new heavy fermions), the experimental
signals of the new physics which can potentially be responsible for EFT Naturalness may be searched
for  in Higgs pair production, for example
in $VV \to hh$ ($V=W,~Z$)\footnote{We note in passing that studies
of $W_L W_L$ scattering have for long been emphasized in theories of strong dynamics.
In contrast, the deviations in $WW \to hh$ that we are suggesting
(as a probe of EFT Naturalness) are caused by weakly coupled physics and not a strongly coupled one.}
due to an
s-channel exchange of an off-shell heavy boson
and in $\bar\psi \psi \to hh$ ($\psi$ is a SM lepton or quark)
due to a t-channel exchange of an off-shell heavy fermion $\Psi$.
Another interesting signature
of the potential role that a heavy fermion
may have in curing the little hierarchy problem of the SM,
is the production of Higgs+quark(jet) or Higgs+lepton,
via its (off-shell) coupling $\Psi^\star \to h \psi$.

\section{Summary}

We have used EFT techniques to analyse naturalness in the SM Higgs sector
in a model independent way,
by calculating the 1-loop contributions to the SM Higgs mass, generated by a set of
higher dimensional effective operators. These higher dimensional
effective operators correspond to exchanges of heavy fermions, scalars and gauge-bosons
in generic underlying renormalizable gauge theories with a typical mass
scale $M \geq \Lambda$, where $\Lambda$ is defined to be the scale
below which these theories are natural.

This framework allows us to find the
conditions under which the 1-loop corrections to $ m_h $  from the new heavy physics
can balance those created by SM loop effects up to the naturalness scale
$\Lambda$, a condition we denote by ``EFT Naturalness''.
Our EFT Naturalness conditions, therefore,
depends on the
heavy particle spectrum and its interactions with the SM Higgs and
it proved to be useful
for acquiring insight regarding the underlying heavy theories that
can address the naturalness problem of the SM, in particular,
for deriving relations among the parameters of the underlying NP theory.








\end{document}